\documentstyle[12pt,epsfig]{article} 
\def\baselinestretch{1.3}
\newcommand{\ba}{\begin{array}}
\newcommand{\ea}{\end{array}}
\newcommand{\bd}{\begin{displaymath}}
\newcommand{\ed}{\end{displaymath}}
\newcommand{\be}{\begin{equation}}
\newcommand{\ee}{\end{equation}}
\newcommand{\bea}{\begin{eqnarray}}
\newcommand{\eea}{\end{eqnarray}}
\newcommand{\sla}[1]{/\!\!\!#1}

\def\q2 {q^2}

\def\r {\rightarrow}
\def\st {\tilde{t} }
\def\sb {\tilde{b} }

\def\rslep {\tilde{e_R}}

\def\snu {\tilde{\nu}}
\def\lslep {\tilde{e_L}}
\def\stau {\tilde{\tau}}
\def\mer {m_{\rslep}}
\def\mmr {m_{\tilde{\mu}_R}}
\def\mml {m_{\tilde{\mu}_L}}
\def\mel {m_{\lslep}}

\def\bt{\begin{table}}
\def\et{\end{table}}

\catcode`@=11 
\def \gsim{\mathrel{\mathpalette\@versim>}}
\def \lsim{\mathrel{\mathpalette\@versim<}}
\def \@versim#1#2{\lower0.4ex\vbox{\baselineskip\z@skip\lineskip\z@skip
     \lineskiplimit\z@\ialign{$\m@th#1\hfil##\hfil$%
     \crcr#2\crcr\sim\crcr}}}
\catcode`@=12 

\begin{document}
\vspace{-0.5cm}
\begin{flushright}
UCRHEP-T506\\
RECAPP-HRI-2011-003
\end{flushright}

\begin{center}

{\large \textbf {Signatures of supersymmetry with non-universal 
Higgs mass at the Large Hadron Collider}}\\[9mm]

Subhaditya Bhattacharya$^{\dagger}$\footnote{subhab@ucr.edu}, 
Sanjoy Biswas$^{\ddagger}$\footnote{sbiswas@hri.res.in},  
Biswarup Mukhopadhyaya$^{\ddagger}$\footnote{biswarup@hri.res.in}\\
 and Mihoko M. Nojiri$^{\ast}$\footnote{nojiri@post.kek.jp}\\

$^{\dagger}${\em Department of Physics and Astronomy\\
            University of California, Riverside, California 92521, USA}\\[4.0mm] 

$^{\ddagger}${\em Regional Centre for Accelerator-based Particle Physics \\
     Harish-Chandra Research Institute\\
 Chhatnag Road, Jhunsi, Allahabad - 211 019, India}\\[4.0mm] 

$^{\ast}${\em Institute for the Physics and Mathematics of the Universe\\
       University of Tokyo, Chiba 277-8583, Japan\\
Theory Group, KEK, 1-1 Oho, Tsukuba, Ibaraki 305-0801, Japan\\
The Graduate University for Advanced Studies (SOKENDAI)\\
       1-1 Oho, Tsukuba, Ibaraki 305-0801, Japan}\\[7mm]


\end{center}

\begin{abstract} 

We discuss large non-universality in the Higgs sector at high scale in 
supersymmetric theories, in the context of the Large Hadron Collider (LHC).  In particular, we note 
that if ${m_{H_u}}^2-{m_{H_d}}^2$ is large and negative ($\simeq 10^6 {\rm ~GeV^2}$) at high scale, 
the lighter slepton mass eigenstates at the electroweak scale are mostly left chiral, in contrast to 
a minimal supergravity (mSUGRA) scenario. We use this feature to distinguish between non-universal
Higgs masses (NUHM) and mSUGRA by two methods. First, we study final states with same-sign ditaus.
We find that an asymmetry parameter reflecting the polarization of the taus provides a notable 
distinction. In addition, we study a charge asymmetry in the jet-lepton
invariant mass distribution, arising from decay chains of left-chiral squarks leading to
leptons of the first two families, which sets apart an NUHM scenario of the above kind.

\vskip 20pt
\end{abstract}

\newpage
\setcounter{footnote}{0}

\def\baselinestretch{1.5}
\section{Introduction}

With the Large Hadron Collider (LHC) already running, one feels closer
than ever to glimpses of physics beyond the standard
model (SM). Supersymmetry (SUSY) \cite{Book, Sally} has always remained an 
attractive hunting ground in this context. The LHC has brought added impetus to
not only the search for SUSY, but also the more ambitious proposal to
identify the overseeing high-scale physics that can lead to typical
low-energy spectra.  Such high-scale physics is often envisioned as
the `organizing principle' behind the plethora of low-scale parameters in the
minimal supersymmetric standard model (MSSM) \cite{Martin1,Djouadi:1998di} to 
be seen at low energy. A frequently adopted approach in this direction is to 
embed MSSM in a minimal supergravity (mSUGRA) scenario \cite{mSUGRA}, where 
all the low-scale parameters can be generated from:
\bea
\nonumber
m_{1/2}, m_0, A_0,\tan\beta~ {\rm and}~ sgn(\mu)
\eea
where $m_{1/2}, m_0, A_0$ are the universal gaugino mass, scalar mass
and trilinear scalar coupling parameters respectively at the high-scale, 
$\tan \beta= \langle H_u \rangle/ \langle H_d \rangle$ is the ratio of the two
Higgs vacuum expectation values and $\mu$ is the SUSY-conserving
Higgsino mass parameter in the MSSM superpotential.

However, the mSUGRA model can be branded over-simplistic, as the
assumption of universality doesn't follow from any known symmetry
principle. For example, gaugino mass non-universality can occur in
supersymmetric Grand Unified Theories (SUSY-GUT) \cite{GUT} with non-trivial
gauge kinetic functions \cite{nonunigaugino,SO10GUT}. Non-universality in the 
scalar sector can
also be motivated from the $SO(10)$ $D$-terms \cite{SO10D}, apart from the
phenomenological requirement to keep CP-violation and flavour changing
neutral currents (FCNC) under control \cite{Misiak:1997ei}. Also, 
in mSUGRA, one assumes that the Higgs mass
parameters have their origin in the same $m_0$ which generate squark
and slepton masses, which is completely {\em ad hoc}.  For example, in
SUSY-GUT theories based on the $SO(10)$ group, sfermions and Higgs
fields belong to different representations and can therefore arise
from independent high-scale mass parameters.  In view of this, one
can, within the SUGRA scenario itself, expect the Higgs mass
parameters to arise from high scale value(s) different from $m_0$.
Thus models with non-universal Higgs mass (NUHM) are of considerable
interest, and their viability in respect of both collider signals and
issues such as the dark matter content of the universe has been
recently investigated \cite{NUHMT,Baer,NUHMP}.

One can incorporate the non-universality in the Higgs sector in two
different ways. In the first kind, one can have both of the soft Higgs
mass parameters originating in a high-scale value $m_0^{'}$ which is
different from $m_0$, the universal high-scale mass for squarks and sleptons.
In other words, one can postulate
$m_{H_u}^2=m_{H_d}^2={m_0^{'}}^2\neq m_0^2$ \cite{Baer}.  On the other hand, it
is also possible to have $H_u$ and $H_d$ evolve down  from
{\em two different high-scale inputs}. In the later case, 
the high-scale SUSY parameters are given by:
\bea
\nonumber
m_{1/2}, m_0, m_{H_u}^2, m_{H_d}^2, A_0,\tan\beta~ {\rm and}~ sgn(\mu)
\eea

The split between the two Higgs squared masses at high scale introduces
additional features in the running of various mass parameters down to
the electroweak scale.  In its most drastic manifestation, such
a situation can give rise to the sneutrino
($\tilde{\nu}$) as the lightest superpartner of standard model
particles. Since a sneutrino dark matter candidate is disfavoured from
available results on direct search, one then has to postulate the
sneutrino(s) to be the next-to-lightest supersymmetric particle(s)
(NLSP), and, for example, gravitino as the lightest supersymmetric
particle (LSP). With this achieved, most of the allowed region of the
NUHM parameter space leads to the right amount of relic density 
\cite{Bhattacharya:2009ij}.

In this work, we propose using the LHC data to distinguish those cases
where the superparticle spectrum in NUHM is most strikingly different
from the usual mSUGRA scenario. As we shall see in the next
section, this happens for a large negative high-scale value of
${m_{H_u}}^2-{m_{H_d}}^2$. It not only leads to a large splitting
between the left and right chiral sleptons, but also leads to the
lighter slepton mass eigenstate of any flavour being dominated by the
left chiral component.

This feature, marking a drastic departure from the expectations in
mSUGRA, can be reflected in the signals of staus through the
polarization of the taus that are produced either in their decay or in
association with them \cite{Nojiri:tau,tau1,Choi:2006mt}. In addition, 
the above hierarchy between
left-and right-chiral sleptons can be probed by studying the spin correlation
of jets and leptons
produced in cascade decays of squarks. This correlation, as we shall see, 
affects
the angular distribution of the lepton in  $\chi^0_2\r l^{\pm}\tilde{l}^{\mp}$,
manifested through certain measurable kinematical variables 
\cite{Richardson:2001df,Barr:2004ze,Nojiri}.

To explain further, the large splitting between the
left-and right-chiral sleptons sometimes yields a hierarchy
where the right-chiral ones become much heavier than not only 
the left-chiral ones but also the low-lying chargino/ second lightest 
neutralinos over a large region of the NUHM parameter space. Thus 
they are hardly produced in collider experiments.  At the same time,
the (dominantly) left-chiral stau and the corresponding sneutrino
being considerably lighter --- even lighter than the lightest
neutralino--- the taus produced in their association are dominantly
left-handed. This is due to the fact that the gauge couplings involved
in the decay are chirality conserving, so long as one has large
gaugino components in the lighter neutralinos and charginos.

The consequences that we focus on are two-fold. First, one notices 
the practically ubiquitous $\tau$ in SUSY signals. Secondly,
the signals often bear the  stamp of left-polarized 
$\tau^-$'s,  in the products of their one-prong decay.
With this in view, we analyze the polarization of the taus produced in
the SUSY cascades in the same-sign di-tau ($SSD\tau$) final states
associated with hard jets and missing transverse energy ($\sla E_T$).
We show how this leads to noticeable differences
between the NUHM and mSUGRA spectra in the LHC environment.

Furthermore, we study the polarization dependence of the angular distribution
of the lepton produced in $\chi^0_2$ decay, which shows up in the charge
asymmetry in the $m_{ql}$ distribution. Though the effect tends to wash out
due to the presence of antisquark decay, nevertheless it can be observed at the
LHC as more squarks are produced than antisquarks.

This paper is organized as follows. We discuss various aspects of the
model under consideration in the following section and identify the
region of the $m_0-m_{1/2}$ parameter space where the lighter stau is dominantly
left-chiral. As we shall see below, this is achieved for large
negative values of $S$.  We choose a few benchmark points for our
collider simulation. Tau-polarization and its
implications are discussed in section 3, while the analysis
revealing the chirality information on sleptons of the first two
families is outlined in section 4.
The numerical results for each of the two analyses mentioned above,
based on a simulation for the 14 TeV run of the LHC,
is presented in section 5. We summarise and conclude in section 6.

\section{Features of the NUHM scenario and our choice of benchmark points}

\subsection{Salient features of the scenario}

We consider the general case of NUHM, having a two-parameter extension of the
mSUGRA scenario, in which the soft SUSY breaking masses $m_{H_u}^2$ and
$m_{H_d}^2$ are inputs at high scale. The most important thing to 
remember  here is that the renormalisation group evolution (RGE) of soft 
scalar masses is in general modified by the presence of a non-zero boundary 
value of the quantity $S$, defined as \cite{Martin1}

\begin{eqnarray}
\hspace{-1.5cm} S &=& m_{H_u}^2-m_{H_d}^2+ 
Tr\left[ {\bf m}_Q^2-{\bf m}_L^2-2{\bf m}_U^2+
{\bf m}_D^2+{\bf m}_E^2 \right] 
\label{eq:S}
\end{eqnarray}

We assume universality in the sfermion masses, so that $S =
m_{H_u}^2-m_{H_d}^2$  is high scale boundary condition.  The
running of soft scalar masses of the third family squarks and 
sleptons are given at the one-loop level by \cite{Martin1}
\begin{eqnarray}
\frac{dm_{Q_3}^2}{dt}&=&{2\over 16\pi^2}\left(-{1\over 15}g_1^2M_1^2-
3g_2^2M_2^2-{16\over 3}g_3^2M_3^2+{1\over 10}g_1^2S+
y_t^2X_t+y_b^2X_b\right)\\
&& \label{rgeq1}\\
\frac{dm_{\st_R}^2}{dt}&=&{2\over 16\pi^2}\left(-{16\over 15}g_1^2M_1^2-
{16\over 3}g_3^2M_3^2-{2\over 5}g_1^2S+2y_t^2X_t\right), \\
\frac{dm_{\sb_R}^2}{dt}&=&{2\over 16\pi^2}\left(-{4\over 15}g_1^2M_1^2-
{16\over 3}g_3^2M_3^2+{1\over 5}g_1^2S+2y_b^2X_b\right), \\
\frac{dm_{L_3}^2}{dt}&=&
{2\over 16\pi^2}\left(-{3\over 5}g_1^2M_1^2-
3g_2^2M_2^2-{3\over 10}g_1^2S+y_\tau^2X_\tau\right), \\
\frac{dm_{\stau_R}^2}{dt}&=&{2\over 16\pi^2}\left(-{12\over 5}g_1^2M_1^2+
{3\over 5}g_1^2S+2y_\tau^2X_\tau\right) . \label{rgeq4}
\end{eqnarray}
{\rm where the notations for squark, slepton and gaugino masses have
their usual meaning, and $t=\log (Q)$,~ $y_{t,b,\tau}$ are the $t$,
$b$ and $\tau$ Yukawa couplings, and}
\begin{eqnarray}
\hspace{-1.7cm} X_t &=& m_{Q_3}^2+m_{\st_R}^2+m_{H_u}^2+A_t^2 ,\\
\hspace{-1.7cm} X_b &=& m_{Q_3}^2+m_{\sb_R}^2+m_{H_d}^2+A_b^2 ,\\
\hspace{-1.7cm} X_\tau &=& m_{L_3}^2+m_{\stau_R}^2+m_{H_d}^2+A_\tau^2 
\label{eq:X_f}
\end{eqnarray}

Mass parameters of the first two family scalars run in a similar
manner, excepting that the Yukawa contributions are vanishingly
small. The main difference in the SUSY particle spectrum with respect
to an mSUGRA scenario is the non-vanishing boundary value of $S$.  If
this boundary value is large in magnitude, the effect on the
spectrum at low scale is naturally a rather pronounced departure from
mSUGRA.  Since the contribution of the term containing $S$ comes with
different factors in the running of left-handed squarks (sleptons) and
right-handed squarks (sleptons), due to different $U(1)$ hypercharge
assignments, one can have large splitting in the left-right sector
within each generation when $|S|$ is substantially large.

One can see from equation (6) and (7) that the effect of non-universal
Higgs mass is rather pronounced in the slepton sector, the primary
reason being that the running masses are not controlled by the strong
sector. The most important difference it makes to the spectrum is that, 
for large negative values of $S$ ( $\mathcal{O}$(TeV)$^2$) ) \cite{NUHMT,Baer}, 
the left-chiral sleptons tend to become considerably lighter than their
right-chiral counterparts.  This is in striking contrast to both
mSUGRA and gauge mediated SUSY breaking (GMSB). An immediate
temptation that the phenomenologist faces, therefore, is to extract
some signature of this `chirality swap' in the lightest sleptons at
the LHC, which may put a distinctive stamp of NUHM on them. This, 
of course, has to be done with the help of leptons that are produced
either in association with the low-lying sleptons or in their decays.
Since the helicity of leptons of the first two families is difficult
to measure in the collider environment, we feel that it is our best
bet to latch on to the copious number of taus arising from SUSY cascades, 
and concentrate on those features of their decay products that
tell us about their helicities.

As has been noted already,  the above effect is seen for large negative $S$. 
Such values of $S$ therefore become the benchmarks for testing the
special features of NUHM, and it is likely that in such condition only
its footprints are noticeable at the LHC.  Thus we examine next the
kinds of spectra ensuing from large negative $S$, and look for their
observable signature.

A large negative $S$ at high scale affects the running of the
third family SU(2) doublet slepton (both the stau and the
tau-sneutrino) masses in the same way as is done by their Yukawa couplings,
thus bringing them down substantially at low energy.  
As a consequence, one can have both of them of the same order as, or
lighter than, the lightest neutralino ($\chi^0_1$). In the latter
situation, the left-chiral tau-sneutrino is lighter than
the corresponding stau due to the SU(2)
breaking D-terms (for $\tan\beta>1$) :

\begin{eqnarray}
m^2_{\tilde{\tau}_L} &=& m^2_L-\cos(2\beta) m^2_Z(\frac{1}{2}-\sin^2{\theta_W})\\
m^2_{\snu_\tau} &=& m^2_L+\cos(2\beta) m^2_Z .\frac{1}{2}
\end{eqnarray}

In such cases, the tau-sneutrino has to be the NLSP, due to its
unsuitability as a dark matter candidate as laid down by direct search
results.  A gravitino, for example, can be envisioned as the LSP and
dark matter candidate in such cases.  The lighter stau mass eigenstate
can in principle also become the NLSP through mixing of the left and
right chiral fields.  However, this happens only in very restricted
regions of the parameter space, as large mixing requires $\tan\beta$
to be on the higher side, a feature that is highly restricted in NUHM
by the requirements of absence of tachyonic states as well
as of electroweak symmetry breaking.


The role of $S$ in the running of $m_{H_u}^2$ and $m_{H_d}^2$ is described by 
\begin{eqnarray}
\frac{dm_{H_u}^2}{dt}&=&\frac{2}{16\pi^2}\left( -{3\over 5}g_1^2M_1^2-
3g_2^2M_2^2+{3\over 10}g_1^2 S+3f_t^2X_t\right),\label{eq:mhu} \\
\frac{dm_{H_d}^2}{dt}&=&\frac{2}{16\pi^2}\left( -{3\over 5}g_1^2M_1^2-
3g_2^2M_2^2-{3\over 10}g_1^2 S+3f_b^2X_b+f_\tau^2X_\tau\right) \label{eq:mhd} ,
\end{eqnarray}

One can see above that a negative $S$ tends to partially cancel the effects of
top quark Yukawa coupling in the running of $m_{H_u}^2$ and make it
positive at low energy. $m_{H_d}^2$, on the other hand, is routinely
rendered positive at low scale due to the gauge interactions, and the
effects of the term proportional to $S$ often fails to make it negative as
one comes down to the electroweak scale.  Consequently, radiative
electroweak symmetry breaking at the right energy requires a negative
value of $m_{H_u}^2$ at high scale.  Of course, one is led to have a
sufficiently large magnitude of $\mu$ to ensure that $m_{H_u}^2 +
\mu^2$ remains positive at high energy.

\subsection{The choice of benchmark points}

As has been already explained, our purpose is to suggest some
observations at the LHC, which will bring out the distinctive
characteristics of the NUHM spectrum. Such distinction is most
pronounced when the chiralities of the low-lying sleptons are reversed
with respect to the corresponding cases in mSUGRA. This, we have
found, is best achieved (and one is indeed optimistic about clear
distinction) when $S$ is large and negative ($\sim 10^6$ GeV$^2$).  For smaller magnitudes
of $S$ ($\le 10^5$ GeV$^2$), the $\stau_L$ component of $\stau_1$
decreases, and the collider signature of this scenario is relatively
less distinct.  With this in view, the region in the parameter space
with more than 90\% of $\stau_L$ in $\stau_1$ has been shown in 
Figure 1. This region offers the best hope for
recognising NUHM if SUSY is detected at the LHC. We have
accordingly chosen some benchmark points for the study reported in the
subsequent sections. Out of the regions answering to our chosen
criterion, we have selected points with three possible mass
hierarchies:
\begin{eqnarray}
m_{\snu_{\tau_L}} < m_{\chi^0_1} < m_{\stau_1} \nonumber\\
m_{\snu_{\tau_L}} < m_{\stau_1} < m_{\chi^0_1} \nonumber\\
m_{\chi^0_1} < m_{\snu_{\tau_L}} < m_{\stau_1} \nonumber
\end{eqnarray}

Our benchmark points (BP) NUHM-1 - NUHM-3 (shown in Table 1) are taken from three 
regions of the parameter space, corresponding to each of the above hierarchies. 
The code SuSpect (version 2.41)\cite{SUSPECT} has been used for this
purpose. Two-loop renormalisation group equations have been used for
running the mass parameters down to low energy, with the default option
(namely, $\sqrt{\tilde{t_1} \tilde{t_2}}$) for the electroweak symmetry
breaking scale. The spectra are consistent with low energy constraints 
\cite{Amsler:2008, constraints} such as those coming from
$b\longrightarrow s\gamma$ and the muon anomalous magnetic moment, and
also with those from LEP-2 limits, such as $m_{\chi^{\pm}_1}>103.5$~GeV,
$m_{\tilde{l}^{\pm}}> 98.8$~GeV and $m_h> 111$~GeV. Electroweak
symmetry breaking in a consistent fashion has been taken as a
necessary condition in the allowed parameter space. For the case with
$ \chi^0_1$ LSP, the requirement of relic density consistent with the
recent data has also been taken into account in choosing the benchmark
point(s) \cite{wmap}.

\vspace{-0.5cm}
\begin{table}[htp]
\begin{center}
\begin{tabular}{||c||c|c|c||}
\hline
\hline
 Benchmark points  & {\bf NUHM-1}&{\bf NUHM-2}&{\bf NUHM-3}\\
\hline
 Input       &$m_0=300$&$m_0=80$&$m_0=300$  \\
 parameters  &$m_{1/2}=300$&$m_{1/2}=460$&$m_{1/2}=280$  \\
             &$\tan\beta=10$ &$\tan\beta=10$&$\tan\beta=7$ \\
\hline
$\mel,\mml$   & 170 & 154 & 154  \\
$\mer,\mmr$   & 552 & 437 & 551  \\
$m_{\snu_{e_L}},m_{\snu_{\mu_L}}$& 151 & 132 & 133  \\
$m_{\snu_{\tau_L}}$& 119 & 106 & 116  \\
$m_{\stau_1}$& 139 & 124 & 137 \\
$m_{\stau_2}$& 537 & 424  & 543  \\
\hline
$m_{\chi^0_1}$ & 120 & 187  & 112 \\
$m_{\chi^0_2}$& 234 & 361 & 216  \\
$m_{\chi^0_3}$& 939 & 982  & 950  \\
$m_{\chi^0_4}$& 944 & 987  & 954   \\
$m_{\chi^{\pm}_1}$& 234 & 361  & 217 \\
$m_{\chi^{\pm}_2}$& 944 & 987  & 955  \\
$m_{\tilde{g}}$& 734 & 1066   & 691   \\
$m_{\tilde{t}_1}$& 645 & 826   &618  \\
$m_{\tilde{t}_2}$& 814 & 1018  &791  \\
$m_{\tilde{d}_L}$& 741 & 986  &706  \\
$m_{\tilde{d}_R}$& 742 & 962  & 710  \\
$m_{\tilde{u}_L}$& 737 & 984  & 701  \\
$m_{\tilde{u}_R}$& 591 & 879 & 549   \\
$m_{h^0}$ &  111  &  114  &  112 \\
\hline
\hline
\end{tabular}\\
\caption {\small \it Proposed benchmark points for the 
study of the NUHM scenario with $m_{H_u}^2=-1.10\times 10^6 {\rm ~GeV}^2$ 
and $m_{H_d}^2=2.78\times 10^6 {\rm ~GeV}^2$. All the mass parameters 
are given in units of GeV. The value of $A_{0}$ is taken to be zero and sign of $\mu$ 
to be positive for all of the benchmark points.} 
\label{tab:1}       
\end{center}
\end{table}




\begin{table}[htp]
\begin{center}
\begin{tabular}{||c||c|c|c||}
\hline
\hline
  Benchmark points     &{\bf mSUGRA-1}&{\bf mSUGRA-2} &{\bf mSUGRA-3} \\
\hline
  Input   &$m_0=80$  &  $m_0=350$  &  $m_0= 300 $ \\
  parameters        &$m_{1/2}=250$  &  $m_{1/2}=300$  & $m_{1/2}= 350 $\\
          &  $\tan\beta=40$  & $\tan\beta=40$  & $\tan\beta= 10$ \\
\hline
$\mel,\mml$   & 389 & 362 &  284  \\
$\mer,\mmr$   &  363 & 322 &  202  \\
$m_{\snu_{e_L}},m_{\snu_{\mu_L}}$ & 381 & 354 &  271 \\
$m_{\snu_{\tau_L}}$ & 353 & 329 & 269 \\
$m_{\stau_1}$ & 283 & 238 &  197 \\
$m_{\stau_2}$ & 377 & 358 &  285  \\
\hline
$m_{\chi^0_1}$ & 99 & 120 &  140  \\
$m_{\chi^0_2}$ & 183 & 224 &   261  \\
$m_{\chi^0_3}$ & 333 & 392  &  455  \\
$m_{\chi^0_4}$ & 352 & 409  &  473  \\
$m_{\chi^{\pm}_1}$ & 182 & 224 & 262   \\
$m_{\chi^{\pm}_2}$ & 353 & 410 & 473    \\
$m_{\tilde{g}}$ & 623 & 726  &  831    \\
$m_{\tilde{t}_1}$  & 464 & 525  &  573   \\
$m_{\tilde{t}_2}$  & 615 & 683   &  764    \\
$m_{\tilde{d}_L}$ & 654 & 720  &  776      \\
$m_{\tilde{d}_R}$  & 636 & 697  &  745     \\
$m_{\tilde{u}_L}$  & 649 & 716  &  771    \\
$m_{\tilde{u}_R}$  & 636 & 698 &   747   \\
$m_{h^0}$  &111 & 112  &  111         \\
\hline
\hline
\end{tabular}\\
\caption {\small \it mSUGRA benchmark points obtained based on similar cross-section
in the same-sign ditau channel (mSUGRA-1 and mSUGRA-2) and in the opposite-sign 
same-flavor dilepton channel (mSUGRA-3). All the mass parameters are given in units of GeV.
The value of $A_{0}$ is taken to be zero and sign of $\mu$ to be positive for all of the benchmark
points.} 
\label{tab:2}       
\end{center}
\end{table}

We have obtained the mSUGRA BP's for comparison with the NUHM points using the
criterion based on similar event rates (within $\pm 30\%$ tolerance) in two different channels. 
For the case where the distinction between these two scenarios is done using tau-polarisation, 
we have compared the event rates in the same-sign ditau ($SSD\tau$) channel 
for choosing our mSUGRA points. For the analysis based upon lepton-charge asymmetry, the event 
rates in the opposite-sign same-flavor dilepton ($OSSFD\ell$) channel have
been compared as a benchmarking criterion. All the three NUHM BP's have been used for
the first case and for the second case, only NUHM-1 and NUHM-3 have been considered, as
the hierarchies mentioned above are not relevant for analysis based on lepton-charge asymmetry.
Thus, we have obtained mSUGRA-1 which corresponds to both NUHM-1 and NUHM-3 following the 
criterion mentioned above in the $SSD\tau$ channel and mSUGRA-3 corresponds to NUHM-3 in the 
$OSSFD\ell$ channel. The benchmark point mSUGRA-2 corresponds to NUHM-2 having similar rates
in the $SSD\tau$ channel. Two of the chosen mSUGRA points (mSUGRA-1 and mSUGRA-2) are approximately 
compatible with the observed relic density.

The values of various SUSY parameters in the chosen points are listed
in Table 1.  One has to further assume in the case of $\snu_{\tau}$
-NLSP and gravitino ($\tilde {G}$) LSP that the decay $\snu_{\tau}\r
\nu_{\tau}\tilde {G}$ does not a lifetime exceeding the age of the universe. 
The gravitino mass has to have accordingly allowed values, as dictated by the 
hidden sector of the overseeing theory \cite{gravitinolsp}.

\begin{figure}[h]
\centerline{\epsfig{file=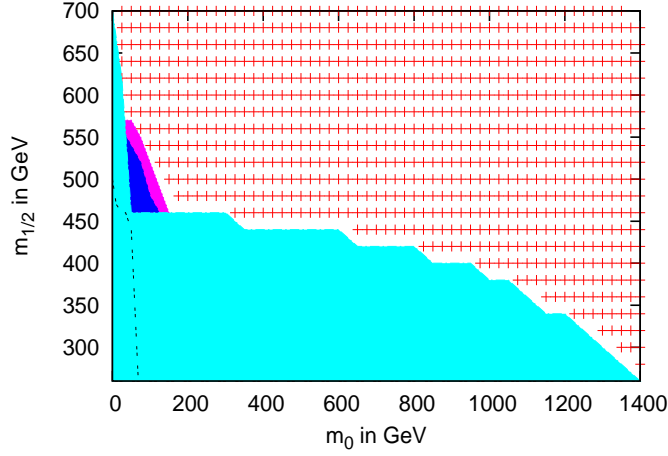,width=6.5cm,height=9.0cm,angle=-90.0}}
\vskip 20pt
\caption{\small \it {The allowed region 
for NUHM in the $m_0-m_{1/2}$ plane for $tan\beta=10$,
$m_{H_u}^2=-1.0\times 10^6 {\rm ~GeV}^2$ , $m_{H_d}^2=2.0\times 10^6 {\rm ~GeV}^2$ and
$A_0=0$. The light blue region is disallowed due to tachyonic stau
and/or non-compliance of electroweak symmetry breaking conditions. The
region on the left of the dashed line is also disallowed by constraints
from $b\rightarrow s \gamma$.  In the region marked by {\small +},
one has $m_{\chi^0_1} < m_{\snu_{\tau_L}} < m_{\stau_1}$, whereas the
pink region corresponds to $m_{\snu_{\tau_L}} < m_{\chi^0_1} <
m_{\stau_1}$ The dark blue region has the hierarchy $m_{\snu_{\tau_L}}
< m_{\stau_1} < m_{\chi^0_1}$.  The lighter stau has 90\% or more of
$\stau_L$ over the entire allowed region.} }
\end{figure}

\section{Tau polarisation}

The signal of left-polarised tau is expected to be a very good discriminator
between scenarios with NUHM and its universal counter part. 
Tau lepton plays a crucial role in the search for new physics. 
In particular, information on the chirality of a tau can be extracted following
some standard procedures. The fact that 
the tau decays within the detector, in
contrast to the electron or the muon, enables us to know about its
chirality from the kinematic distribution of the decay products.  In
the massless limit where the tau is boosted in the laboratory frame,
tau decay products are nearly collinear with the parent tau. In this
limit, hadronic tau decays produce narrow jets of low multiplicity, to
be identified as tau-jets. From the angle of polarisation studies, it
is most cost-effective to work with the one-prong hadronic decay modes
of the tau, which comprise 80\% of its hadronic decay width and about
50\% of its total decay width. The main channels here are:

\bea
& \tau^{-} \rightarrow \pi^{-}\nu_{\tau} & \nonumber \\
&\tau^{-} \rightarrow (\rho^{-}\nu_{\tau}) & \rightarrow \pi^{-}\pi^0\nu_{\tau} 
\nonumber \\
&\tau^{-} \rightarrow (a_1^{-}\nu_{\tau}) &\rightarrow \pi^{-}\pi^0\pi^0\nu_{\tau} 
\nonumber
\eea

\noindent
where we shall often denote both the $\rho^-$ and the $a_1^-$
by $v$.

The first step in the extraction of polarisation information is to
express some differential decay distributions of the $\tau^-$
in the laboratory frame.  Let the polarisation information be 
denoted by $P_\tau$, where   $P_\tau = \pm 1$  correspond
to taus with positive and negative helicity.
Next, it is worthwhile to examine 
the laboratory frame variable $z$,  defined as
$z=E_{\pi,v}/E_{\tau}$,  the fraction of the tau energy
carried by the product meson. 
This variable can be related to  $\theta$,  the angle
between the direction of motion of the outgoing $\pi^-$ or $v^-$ 
and the axis of polarisation of the tau, which is taken to be along the
direction of the tau momentum in the laboratory
frame. In the limit $E_{\tau}>>m_{\tau}$,
\be
\cos\theta=\frac{2z-1-c^2}{1-c^2}
\ee  
where $c={m_v / m_{\tau}}$.  The expression for the case where the tau
decays to the pion and a $\nu_\tau$ is obtained by setting
$m_v = 0$ above.



The decay distributions in $z$ for a $\tau^-$ in the
laboratory frame are given by \cite{tau2}

\bea
{1 \over \Gamma_\pi} {d\Gamma_\pi \over dz} = [1+P_\tau(2z-1)]
\label{z1}
\eea

\bea
\nonumber {1 \over \Gamma_v} {d\Gamma_{v_L} \over dz} =
{m^2_\tau m^2_v \over (m^2_{\tau}-m^2_v)(m^2_\tau + 2m^2_v)} [{m^2_{\tau}\over m^2_v}\sin^2\omega+1+\cos^2\omega
+ P_\tau \cos\theta \\
({m^2_{\tau}\over m^2_v}\sin^2\omega-{m_{\tau}\over m_v}\sin2\omega\tan\theta-1-\cos^2\omega)], 
\label{z2} \\[2mm]
\nonumber {1 \over \Gamma_v} {d\Gamma_{v_T} \over dz} =
{m^2_\tau m^2_v \over (m^2_{\tau}-m^2_v)(m^2_\tau + 2m^2_v)} [{m^2_{\tau}\over m^2_v}\cos^2\omega+\sin^2\omega
+ P_\tau \cos\theta \\ 
({m^2_{\tau}\over m^2_v}\cos^2\omega+{m_{\tau}\over m_v}\sin2\omega\tan\theta-\sin^2\omega)] 
\label{z3}
\eea

\noindent
where 
\bea
\cos\omega=\frac{(m^2_{\tau}-m^2_v)+(m^2_{\tau}+m^2_v)\cos\theta}
{(m^2_{\tau}+m^2_v)+(m^2_{\tau}-m^2_v)\cos\theta}
\eea

In the experiment, one looks for hard jets from the tau,
which corresponds to large values of $z$.   
A close inspection of Equations (\ref{z1}),(\ref{z2}),(\ref{z3}) shows that
the energy distribution of the decay products from the decay of $\tau^{-}_{L}$
($P_{\tau}=-1$) are in significant contrast to that from 
$\tau^{-}_{R}$ ($P_{\tau}=+1$) . When $P_{\tau}=+1$, the hard
$\tau$-jet consist largely of either a single pion or longitudinally
polarised vector mesons ($v_L$). For $P_{\tau}=-1$,  on the contrary,
the hard $\tau$-jet mostly comprises  transversely polarised vector mesons only
($v_T$). This conclusion becomes almost self-evident in,  for example,
the extreme case of collinearity, with $\cos\omega = 1$ and $\sin\omega = 0$.

It should, however, be remembered that the quantity $z$ is not amenable to
actual measurement in the detector, and therefore the distinctions  pointed
out above are still somewhat theoretical in nature. It is therefore necessary
to translate the distinction in terms of measurable quantities.  The energy
distribution among the  pions arising from the decay of the $\rho^-$ and
$a_1^-$ offer such a variable.  It is the variable $R=E_{\pi}/E_{\rho}$,  the fraction 
of the energy of $v$ carried by the charged pion.  For the case where
the $\rho^-$ is produced in $\tau^-$ decay,  the distribution
in $R$ in the laboratory frame is given by \cite{tau2}

\bea
\frac{d\Gamma(\rho_T\rightarrow 2\pi)}{dR} &\sim& 2R(1-R)-
{2m^2_{\pi}\over m^2_{\rho}}\\ 
\frac{d\Gamma(\rho_L\rightarrow 2\pi)}{dR} &\sim& (2R-1)^2
\eea

The distribution for $a_1^-$ is more complicated but has similar qualitative
features.  The reader is referred to \cite{Raychaudhuri:1995kv} for
the detailed expressions. 
The broad indication is that transversely polarised vector mesons favour even
sharing of its momentum among the decay pions whereas longitudinally
polarised ones favour uneven sharing of momentum among its
decay products. Since the polarisation of the parent tau governs the 
level of polarisation of either type in the vector mesons $v$,  the distribution
in the variable $R$ therefore is a reflection of the helicity of the tau whose
signal one is concerned with.

Obviously, one always has $R$ = 1 when the tau decays as 
$\tau^{-} \rightarrow \pi^{-}\nu_{\tau}$.  What one must utilise, therefore,
is the difference in R-distributions between the cases with $v_T$ and $v_L$.
When the decaying tau has $p_\tau$ = +1,  one should mostly have
$v_L$ in the hard jets, in addition to the inconsequential single pions,
giving its characteristic distribution on $R$. A contrast can be seen in the 
decay of a tau with $p_\tau$ = -1, where the hard tau-jets can be expected
to be largely $v_T$,  with a different distribution in $R$.

Hence, one can use the charged-pion spectra arising from the two-stage decays 
\bea
&\tau^{-} \rightarrow (\rho^{-}\nu_{\tau}) & \rightarrow
\pi^{-}\pi^0\nu_{\tau} \nonumber \\ &\tau^{-} \rightarrow
(a_1^{-}\nu_{\tau}) &\rightarrow \pi^{-}\pi^0\pi^0\nu_{\tau} \nonumber
\eea
to probe  the polarisation of the parent tau. We utilise
this possibility to identify the NUHM spectrum in cases the
low-lying stau is of left chirality, which attaches similar chirality
(same as helicity at high energy)  to the taus either arising from
stau-decay or produced in association with it.  With this in view, 
we have selected tau-jets in our simulation with $p_T> ~40 GeV$ and $|\eta|<2.5$,  
assuming a tau jet identification efficiency of 50\%, with a fake tau jet rejection
factor of 100 \cite{tau1}.

\section{Lepton charge asymmetry}

Another discriminator which is sensitive to the mass hierarchy between
the right-and left-chiral sleptons is the charge asymmetry in the
jet-lepton invariant mass distribution \cite{Richardson:2001df,
Barr:2004ze,Nojiri}. In the NUHM scenario (with
large negative $S$), the lighter slepton mass eigenstate is dominated
by the left-chiral component ($\tilde{l}_1\sim\tilde{l}_L$). Hence,
for $m_{\tilde{l}_1}<m_{\chi^0_2}$ (a criterion mostly satisfied by
the `extreme' NUHM scenario considered by us), the leptons produced in
the decay of $\chi^0_2$ will be mostly left-handed. In the usual
mSUGRA scenario, on the other hand, one expects the leptons to be
mostly right-handed as the lighter slepton mass eigenstate is
dominantly right-chiral ($\tilde{l}_1\sim\tilde{l}_R$) and the decay
proceeds via the Bino component of $\chi^0_2$.

This feature can be exploited to unmask NUHM by studying the charge
asymmetry in the lepton-jet invariant mass ($m_{jl_1}$) distribution
produced in the squark decay chains, where $l_1$ stands for the lepton
reproduced in $\chi^0_2$ decay. We shall consider sleptons
of the first two generations only, for which left-right mixing is negligible,
and the coupling of the leptons to the Higgsino components of a neutralino
is also very small.

In this section we describe the spin correlation in the following decay chain 
\bea
\tilde{q_L}\r q\chi^0_2\r ql_1^{\pm}\tilde{l^{\mp}}\r ql_1^{\pm}l_2^{\mp}\chi^0_1
\eea
where $l_2$ denotes the lepton produced in the subsequent step of the cascade.
Due to the chiral structure of the squark-quark-neutralino coupling,
the quark produced in the squark decay will be left-handed in the
massless limit. The $\chi^0_2$ produced in $\tilde{q}_L$ decay is also
polarized having the same helicity as that of the quark as they
are produced from the decay of a scalar.

In the rest frame of the squark produced in the initial hard scattering, 
a negatively charged lepton produced in the subsequent
decay of the $\chi^0_2$ will appear back-to-back or in the same direction
as that of the quark depending on whether the slepton is left-chiral
or right-chiral.\footnote{The other inputs that go into this argument
are (a) The $\chi^0_2$ produced in squark decay is sufficiently
boosted, and (b) the $\chi^0_2$ decays largely in the s-wave.} Exactly 
the opposite directional preferences hold for a
(positively charged) antilepton vis-a-vis the quark produced in the
chain. Therefore, we expect an asymmetry between the distributions
$m_{jl^-_1}$ and $m_{jl^+_1}$. This can be utilised to define the
following asymmetry parameter:

\bea
A_i=\frac{N_i(m_{jl^+_1})-N_i(m_{jl^-_1})}{N_i(m_{jl^+_1})+N_i(m_{jl^-_1})}
\eea
where $i$ stands for the {\it ith} bin. A measurement $A_i$ should
thus yield information on the chirality of the low-lying slepton
produced in the chain.

However, there are some experimental difficulties involved in the measurement 
of such an asymmetry-- 

\begin{enumerate}
\item In the decay of a $\tilde{q}^*_L$, the asymmetry in the lepton-jet
invariant mass distribution has a sign opposite to that of the
corresponding $\tilde{q}_L$. This is because the left antisquark
decays via gaugino coupling into a right-handed antiquark. Since jets
initiated by a quark or an anti-quark are indistinguishable, it is
impossible to disentangle the squark and antisquark production
channels. However, the LHC is a $pp$ machine where more squarks are
produced than anti-sqaurks, a significant `net' charge asymmetry in
the $m_{jl_1}$ distribution can finally survive. All one needs in
order to measure this charge asymmetry is a substantial excess in the
production of $\tilde{q}^{(*)}\tilde{g}$ and
$\tilde{q}^{(*)}\tilde{q}^{(*)}$ over pairs containing squarks and
antisquarks, and also gluino pairs.

\item In an experiment, it is not always possible to distinguish 
between the lepton ($l_1$) out of a $\chi^0_2$ and the lepton ($l_2$)
coming from slepton decay. We have taken the invariant mass
distribution using the harder of the two leptons, a role in which
$l_1$ fits in most of the time.
\end{enumerate}

In NUHM, one expects negative charge asymmetries, whereas in
the usual mSUGRA scenario they are expected to be positive,
especially in the high invariant mass bins. However, in
mSUGRA, depending on the mass hierarchy, the leptons 
produced in $\chi^0_2$ decay can also be dominantly left-handed if
$m_{\tilde{l}_1}<m_{\tilde{l}_2}<m_{\chi^0_2}$, as the diagonal
component ($(U_N)_{22}$) of the neutralino mixing matrix wins over
$(U_N)_{21}$. In that case, one would expect a dip in the asymmetry
distribution at a lower value of $m_{jl}$ and a peaking
behaviour at the higher end. This is expected because the splitting
between $m_{\chi^0_2}$ and $\tilde{l}_L$ is smaller than that between
$m_{\chi^0_2}$ and $\tilde{l}_R$. One can use this feature
to separate an mSUGRA-type scenario.

\section{Collider simulation and numerical results}

We have simulated events for $\sqrt{s}$ = 14 TeV, including
initial-and final-state radiation, multiple scattering etc. We have
used parton distribution functions CTEQL6L1 \cite{Lai:1999wy} for our analysis, with the
renormalisation and factorisation scales set at the average mass of
the final state particles.

\subsection{Simulation strategy: ditau final states}

To study the polarisation of the tau in SUSY cascade for both NUHM and
mSUGRA scenario we have used the code TAUOLA (version 2.9) \cite{tauola} interfaced
with the event generator PYTHIA (version 6.4.16) \cite{PYTHIA}. The spectrum
has been generated using SuSpect (version 2.41) \cite{SUSPECT}. TAUOLA has been
suitably modified to incorporate the probability of producing left-or 
right-handed tau in the decay of SUSY particles.  For cases where
the $\tilde{\nu_\tau}$ and/or the $\stau$ is lighter than the lightest
neutralino, decay branching fractions of the lightest neutralino have
been calculated using SDECAY (version 1.3b) \cite{SDECAY} and fed into
Pythia. The finite detector resolutions have been taken into account
following the specifications listed, for example, in \cite{Biswas1}.

The final state that we have considered is a pair of same-sign ditaus
(SSD$\tau$), together with at least three hard central jets and large
missing $E_T$.  Same-sign ditaus are preferred because they are less
beset with SM backgrounds. We consider events where the taus have
one-prong hadronic decays.

The following cuts have been imposed on each event--

\begin{itemize}
\item  $p_T >$  40 GeV, $|\eta| <$  2.5 for each tau jet.
\item $p_T >$  100, 100, 50 GeV, $|\eta| <$  2.5 for
the three associated jets, in decreasing order of hardness. 
\item $\sla{E}_T >$ 150 GeV.
\end{itemize}

It should be reiterated that our main purpose is to
obtain the observable difference between the NUHM scenario
under consideration and an mSUGRA scenario.
Situations in mSUGRA leading to tau-rich final states are most
likely to fake NUHM phenomenology. Therefore, we have followed the criteria
already mentioned in section 2.2, and isolated
the regions where the total rate of $SSD\tau+\ge 3~jets+\sla{E}_T$
is within $\pm$30\% of the rate predicted for corresponding NUHM benchmark point.

\subsection{Simulation strategy: lepton charge asymmetry}

The charge asymmetry in the lepton-jet invariant mass distribution has
been studied using the event generator HERWIG (version 6.5) \cite{Herwig}
which takes into account the spin correlation in SUSY
cascades. Spectra have been generated using ISAJET (version 7.78)
\cite{isajet} and the input parameters have been tuned in such a way that the
spectrum generated is similar to that produced by SuSpect. A fast
detector simulation has been done using AcerDET (version 1.0) \cite{AcerDET}
for reconstructing the isolated leptons, jets and $\sla{E}_T$, which
also takes into account the finite detector resolution of the visible
momenta.

The final state under consideration is consists of a pair of isolated
leptons of opposite charge and same flavor (OSSF) with more than three
jets and missing $E_T$, i.e., $e^+e^-+\mu^+\mu^-+\ge3~
jets~+\sla{E}_T$.

The preselection cuts \cite{CMS1,TDR} imposed in this case are the following--

\begin{itemize}
\item  $p_{T_{l_1}} >$  20 GeV and $p_{T_{l_2}} >$  10 GeV, $|\eta| <$  2.5 
for the two leptons.
\item $p_T >$  100, 50, 50 GeV, $|\eta| <$  2.5 for
the three associated jets, in decreasing order of hardness.
\item $M_{eff}>$ 600 GeV where, $M_{eff}=\sla{E}_T+\Sigma |\vec{p}_T|$\\
where, the summation is taken over all visible particles.

\item $\sla{E}_T >$0.2$M_{eff}$
\end{itemize}

The SUSY backgrounds come mainly from two independent $\chi^{\pm}_1$
decay. One can eliminate this by taking the flavor subtracted combination
$e^+e^-+\mu^+\mu^--e^{\pm}\mu^{\mp}$ and this cancels out the
background contribution from the charginos up to statistical
fluctuations. The Standard Model background, already small
after imposing the above cuts, undergo further suppression in this process \cite{TDR}.

The leptons are combined with each of the two hardest jets and, for identifying the
desired decay chain, the combination for which the $jl^+l^-$ invariant
mass is smaller has been selected . The $m_{jl^{\pm}}$ distribution
for this subsample, for both the hard and soft lepton have been
calculated. Depending on the mass splitting between the neutralino and
slepton one of these leptons will be dominated by the 'correct'
lepton, i.e., the one adjacent to the quark in the decay chain and
will give the desired charge asymmetry in the jet-lepton invariant
mass distribution.


\subsection{Numerical results}

\vspace{0.5 cm}
{\noindent
{\em {\underline{Ditau final states:}}}
\vspace{0.5 cm}

We first present the numerical results of our analysis
using of the polarisation properties of the tau. In Table
2, we have tabulated the event rates for all the NUHM and the
potentially faking 
mSUGRA points for the $SSD\tau$-channel. Event rates have been
predicted for an integrated luminosity of $100~fb^{-1}$. After
applying all the cuts to suppress the SM background, one has similar
event rates for both the NUHM and corresponding mSUGRA points, which
is not surprising because we have identified the mSUGRA points
following the criterion of similar event rate.

\begin{table}[h]
\begin{tabular}{||c|c|c|c|c||}
\hline
\hline
 {\bf NUHM-1}&{\bf NUHM-2}&{\bf NUHM-3}&{\bf mSUGRA-1}&{\bf mSUGRA-2}\\
\hline

31 & 51 & 28 & 41 & 46 \\

\hline
\hline
\end{tabular}\\
\caption {\small \it Number of events in the $SSD\tau$ channel at an integrated 
luminosity of $100 ~fb^{-1}$ after applying the cuts listed in Section 5.1, in addition 
to a cut on the $R$ variable ($R>0.2$) for all of our benchmark points .} 
\label{tab:2}       
\end{table}

For the benchmark point NUHM-1, $\snu_{\tau_L}$ is the LSP, and the
lighter $\stau_1$ is dominantly left-chiral. Taus are mainly produced
in the decay of $\chi^0_2\r \tau\stau$(20.5\%), $\chi^{\pm}_1\r
\tau\snu_{\tau_L}$(26.5\%) and $\stau\r\tau\chi^0_1$(100\%) and 
therefore the taus are mostly left-handed. The contributions from
$\chi^0_3$, $\chi^0_4$, $\chi^{\pm}_2$ and $\stau_2$ are negligible as
they are heavier in the spectrum. For NUHM-2 we similarly have lighter
stau mass eigenstate dominated by the left-chiral component but here
the mass hierarchy between the $\stau_1$ and the $\chi^0_1$ is
opposite to that of NUHM-1, i.e. $m_{\stau}< m_{\chi^0_1}$. At this benchmark point,
$\chi^0_1$ decays into $\tilde{l}l$ pair as well as $\snu\nu$ pair
including the third generation. The mass difference between the
lighter stau and tau-sneutrino is less than $m_W$, hence the decay
proceed mainly via the two body decay mode
$\stau^{\pm}_1\r\snu^{*}_{\tau}\pi^{\pm}$ and the three body decay
$\stau^{\pm}_1\r\snu^{(*)}_{\tau}l^{\pm}\nu^{(-)}$. However, final states
with higher pion multiplicities also have non-zero branching fractions, 
but we have not taken into account these modes, as they
do not change our conclusion. In NUHM-3, the LSP is the lightest
neutralino, however we still have a light enough $\snu_{\tau_L}$. 
The lighter $\stau_1$, of course, dominantly left-chiral here. The taus
produced in SUSY cascade therefore are mostly left-chiral for all the
NUHM points.  The corresponding R-distributions (taking into account the SM contributions) 
for the respective
benchmark points have been shown in Figure 2. Thus the distinction criterion
set down by us is seen to survive the washouts caused by various extraneous
SUSY cascades.

\begin{figure}[hp]
\begin{center}

\centerline{\epsfig{file=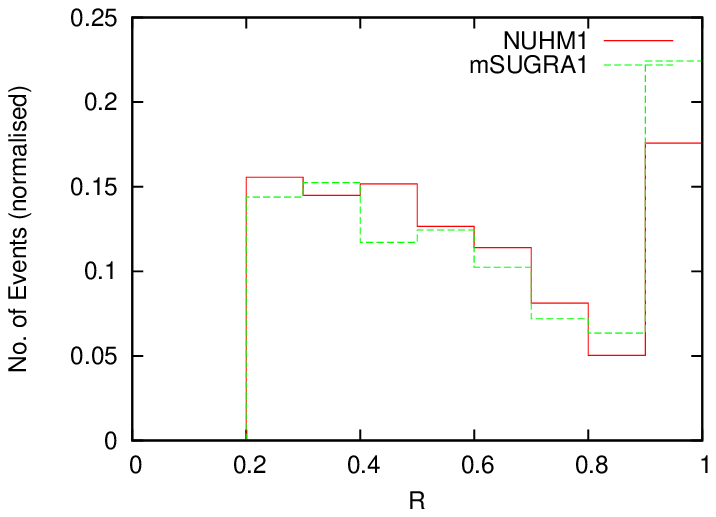,width=8.0cm,height=7.0cm,angle=-0}
\hskip 10pt \epsfig{file=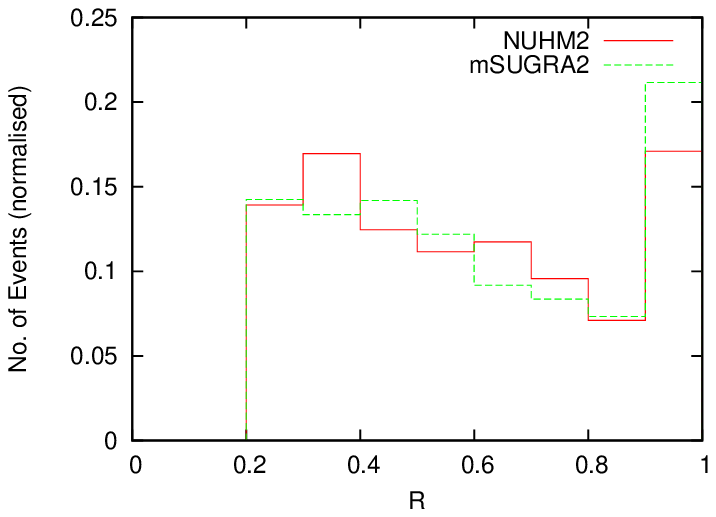,width=8.0cm,height=7.0cm,angle=-0}}
\vskip 15pt
\centerline{\epsfig{file=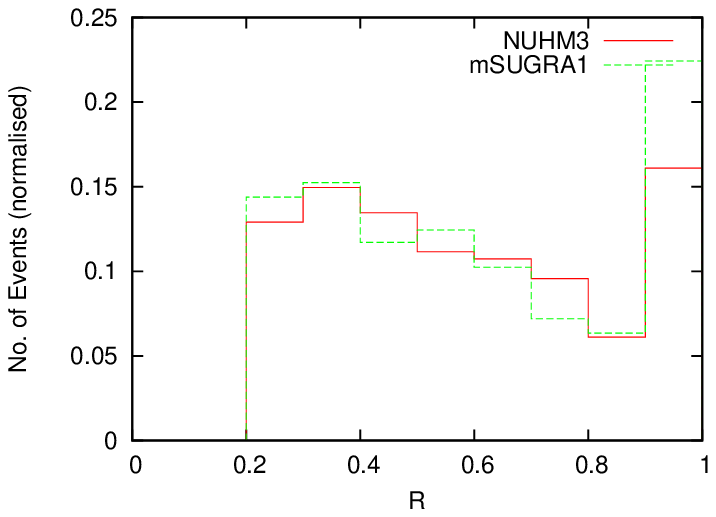,width=8.0cm,height=7.0cm,angle=-0}}
\vskip 10pt
\caption{\small \it {R distribution (defined as $R=E_{\pi^{-}}/E_{\tau_j}$) 
for NUHM scenarios and corresponding mSUGRA points. A cut $R>0.2$ has been
applied in each of these distribution.}} 
\end{center}
\end{figure}


\begin{table}[h]
\begin{tabular}{||c||c|c|c|c|c||}
\hline
\hline
       & {\bf NUHM-1}&{\bf NUHM-2}&{\bf NUHM-3}&{\bf mSUGRA-1}&{\bf mSUGRA-2}\\
\hline

${\mathcal O}_1(R<0.8)$ & 0.77 & 0.74 & 0.76 & 0.71 & 0.72 \\
\hline
${\mathcal O}_2(R>0.8)$ & 0.23 & 0.22 & 0.24 & 0.29 & 0.28 \\
\hline
$r=\frac{{\mathcal O}_1(R<0.8)}{{\mathcal O}_2(R<0.8)}$ & 3.35 & 3.36 & 3.17 & 2.45 & 2.57 \\
\hline
\hline
\end{tabular}\\
\caption {\small \it The ratio $r$ for NUHM and corresponding mSUGRA scenario. } 
\label{tab:3}       
\end{table}

In the corresponding mSUGRA benchmark points (mSUGRA-1 and mSUGRA-2), 
the lighter stau is
dominantly right-chiral. However both the stau are heavier than the
second lightest neutralino and lightest chargino. Taus are produced
mainly via the decay of $W$ and $Z$ produced in the decay of
$\chi^0_{3,4}\r (\chi^{\pm}_1W^{\mp}), (\chi^0_2h/Z)$, $\chi^{\pm}_2\r
(\chi^0_{1,2}W^{\pm}),(\chi^{\pm}_1h/Z)$ and $\chi^{\pm}_1\r
\chi^0_1W^{\pm}$. Hence the contributions to $SSD\tau$ channel come
from two same sign W-decay produced in SUSY cascade or one from
W-decay and one in Z decay, when one of the two tau out of a $Z$-decay
is identified. Therefore, taus are mostly left-handed, with some right-handed
admixtures. This shows up in the R-distribution for the
corresponding mSUGRA points with a slight departure from that of the
NUHM points.

Designating the total number of events for $0.2<R<0.8$ and $R>0.8$ by
${\mathcal O}_1$ and ${\mathcal O}_2$ respectively, we find that the ratio
$r={\mathcal O}_1/{\mathcal O}_2$ is a rather effective discriminator
between NUHM and a corresponding mSUGRA scenario yielding a similar
number of same-sign ditau events. The values of this ratio for all the
cases are listed in Table 3. For the NUHM points, this ratio turn out
to be consistently larger than the corresponding mSUGRA points, which
is expected from the R-distribution given in Figure 2.

\vspace{0.5 cm}
{\noindent
{\em {\underline{Lepton charge asymmetry:}}}
\vspace{0.5 cm}

The results of charge asymmetry in the jet-lepton invariant mass
distribution have been shown in Figure 3 and 4. For both the NUHM-1 and
NUHM-3 benchmark points, gluinos and left-chiral squarks have closely
spaced masses. Therefore, the hard jets are produced either in the decay
$\tilde{q}_L\r q\chi^0_2$ or in $\tilde{q}_R\r q\chi^0_1$, but not in
whichever is allowed between  $\tilde{g}\r q\tilde{q}_{L,R}$ or $\tilde{q}_{L,R}
\r\tilde{g}q$. This is due to 
small mass splitting between them; even if the gluinos are lighter
than the left-chiral squarks, the decay chain $\tilde{q}_L\r
q\chi^0_2\r ql_1^{\pm} \tilde{l^{\mp}}\r ql_1^{\pm}l_2^{\mp}\chi^0_1$
is still the dominant source of the opposite sign
same flavor dilepton signal, as the decay branching ratio of
$\tilde{q}_L\r q\tilde{g}$ is very small ($\simeq 2\%$ or less) due to
phase-space suppression. The branching fraction for $\tilde{q}_L\r
q\chi^0_2$ is $\simeq 32\%$ and subsequently $\chi^0_2$ decays into a
$\tilde{l}^{\pm}l^{\mp}$ pair with a decay branching fraction ranging
from 21\%-29\%, while the sleptons decay into a lepton and the
lightest neutralino with 100\% branching ratio.

\begin{figure}[htbp]
\centerline{\epsfig{file=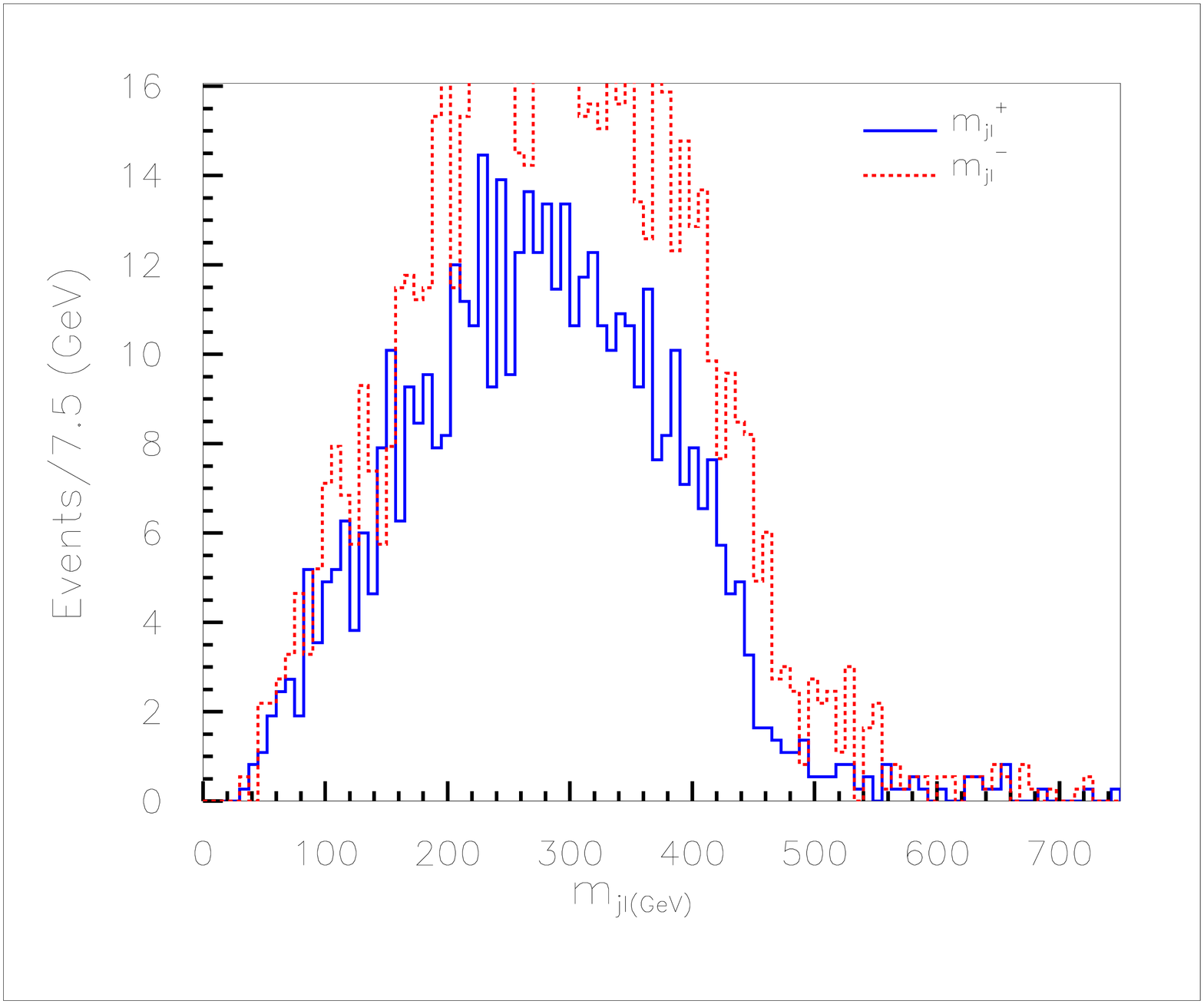,width=7.0cm,height=6.0cm,angle=-0}
\hskip 20pt \epsfig{file=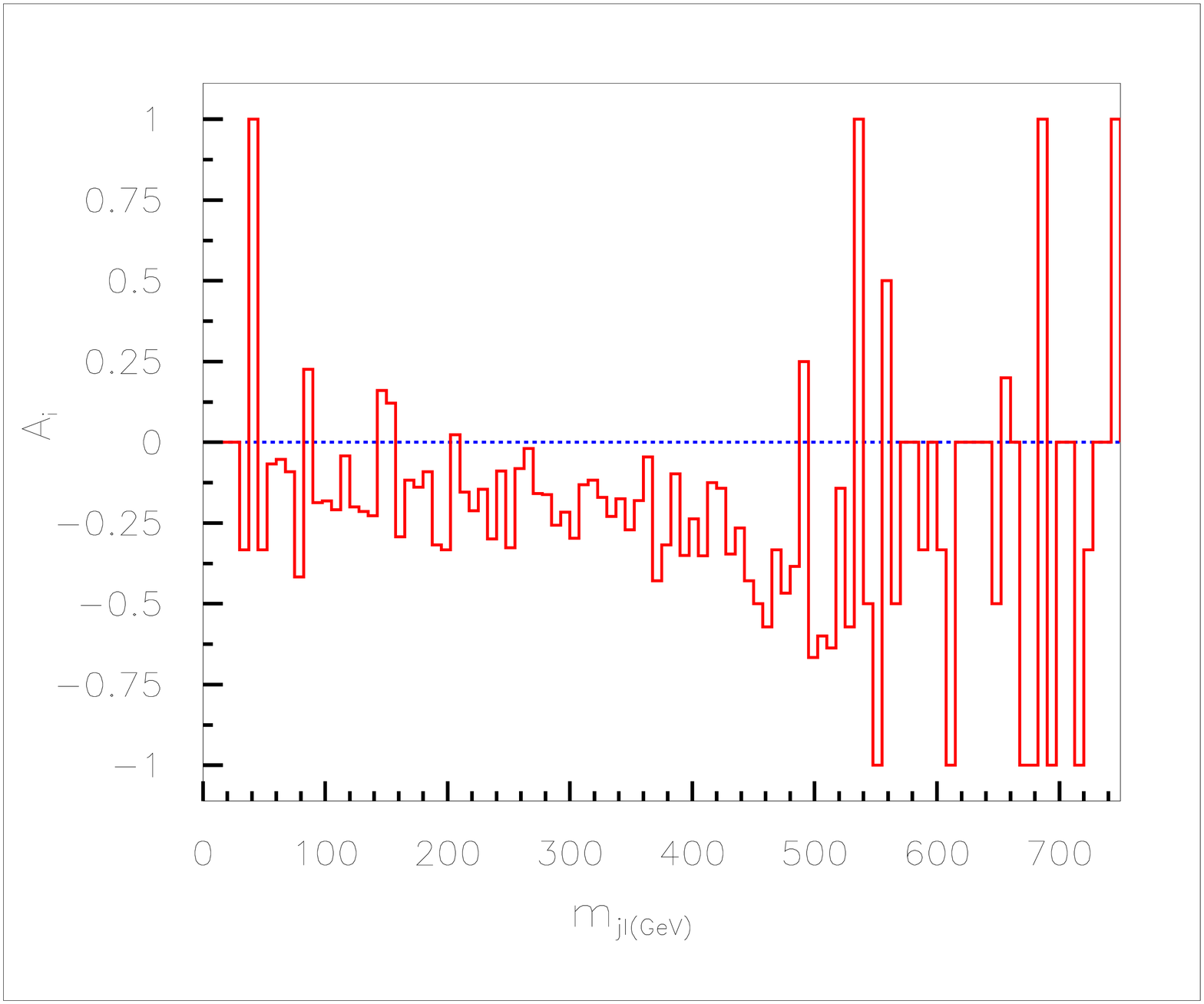,width=7.0cm,height=6.0cm,angle=-0}}
\vskip 15pt
\centerline{\epsfig{file=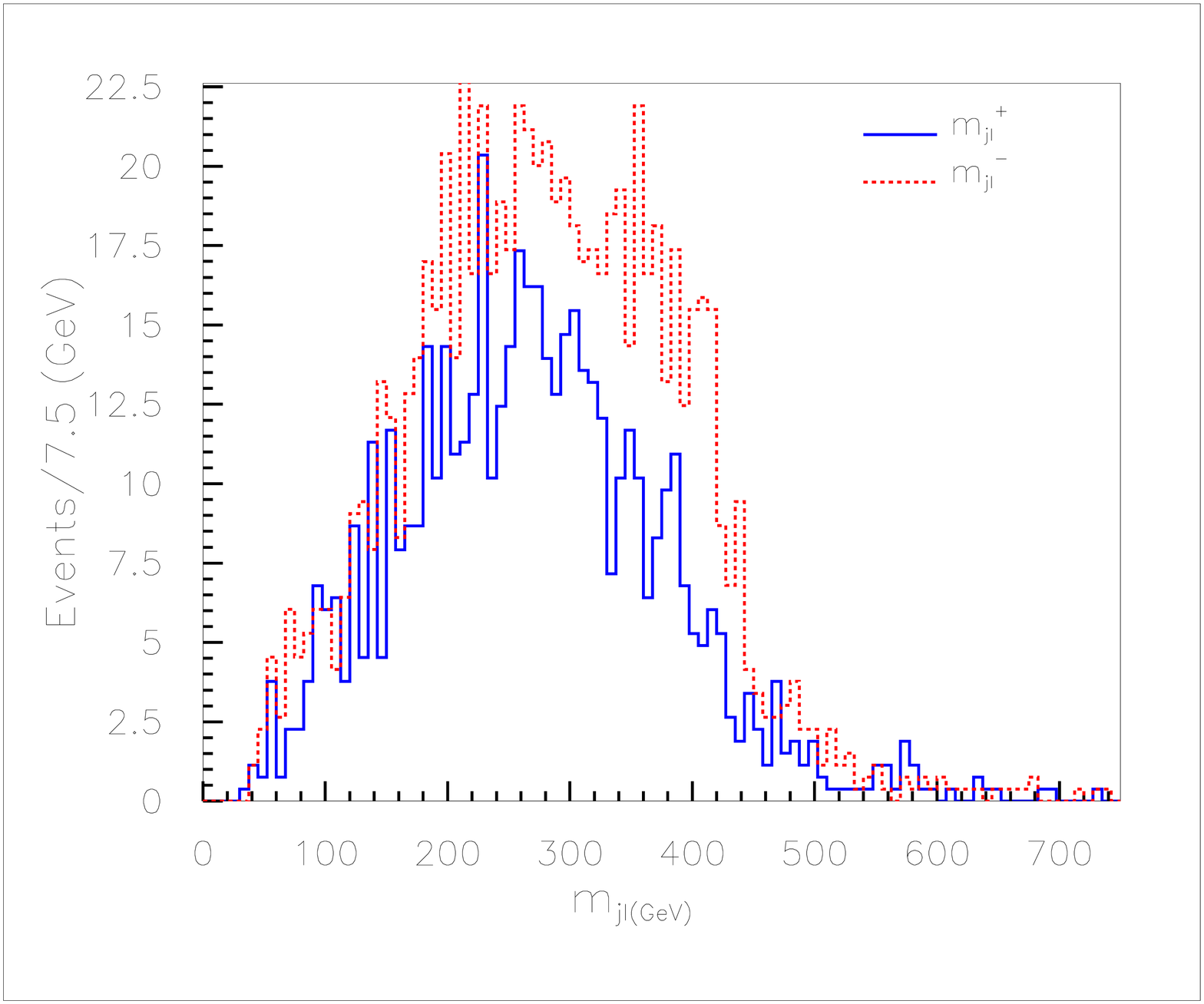,width=7.0cm,height=6.0cm,angle=-0}
\hskip 20pt \epsfig{file=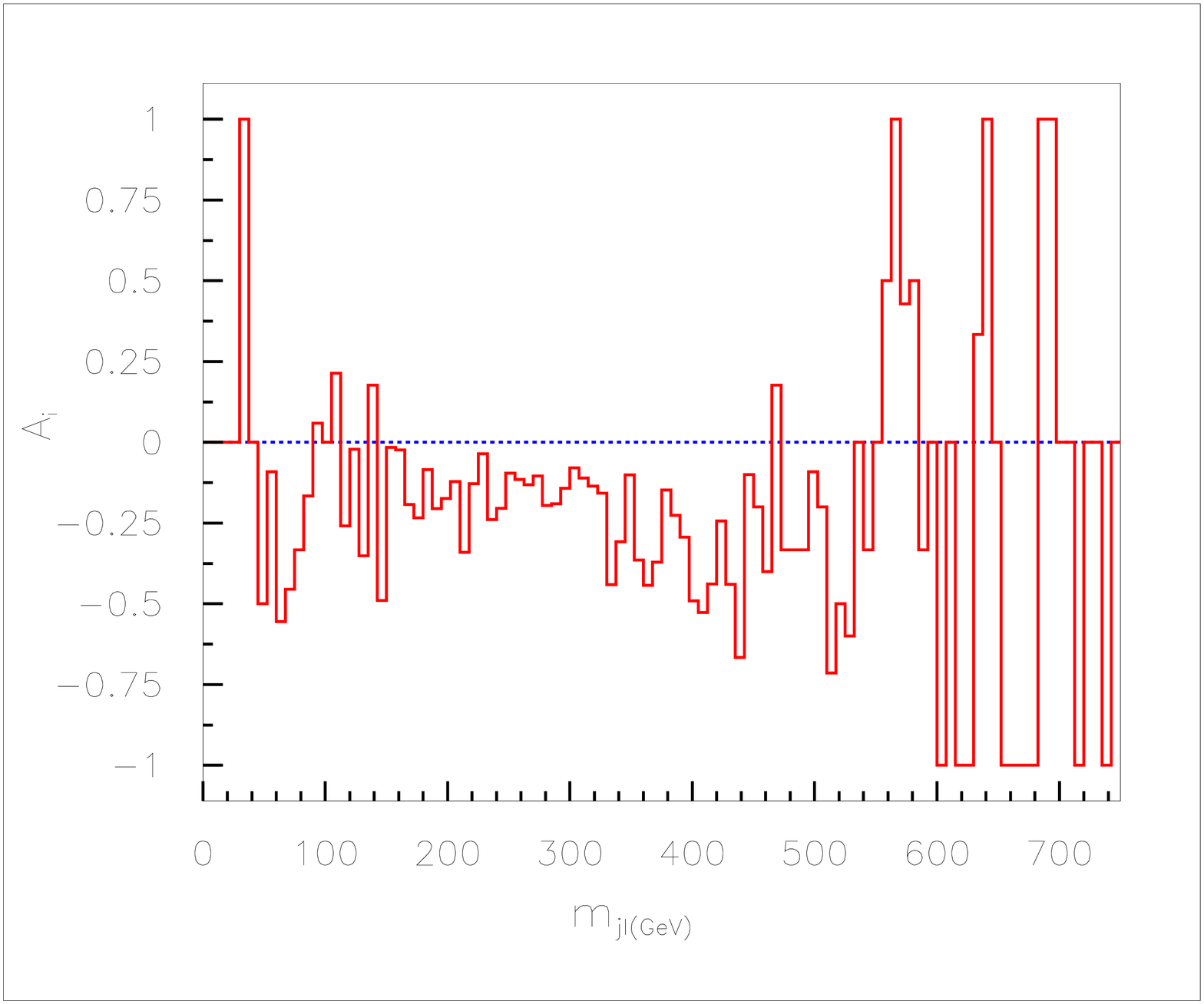,width=7.0cm,height=6.0cm,angle=-0}}
\vskip 20pt
\caption{\small \it {a) $m_{jl}$ and b) $A_i$ vs $m_{jl}$ distribution 
for NUHM BP1 (top) and NUHM BP3 (bottom). The event rates
are predicted at an integrated luminosity of $10 ~fb^{-1}$.}} 
\end{figure}


\begin{figure}[htbp]
\centerline{\epsfig{file=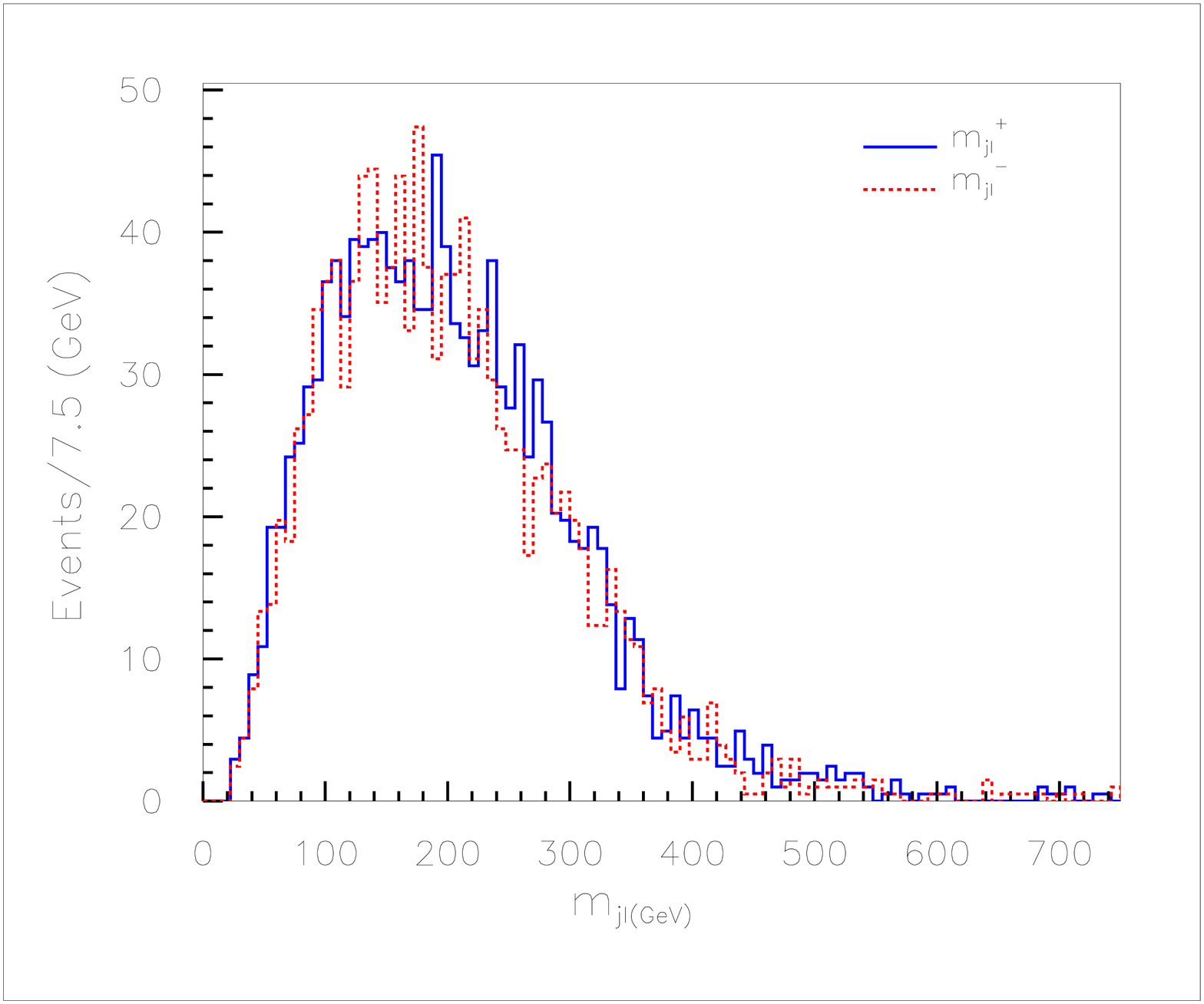,width=7.0cm,height=6.0cm,angle=-0}
\hskip 20pt \epsfig{file=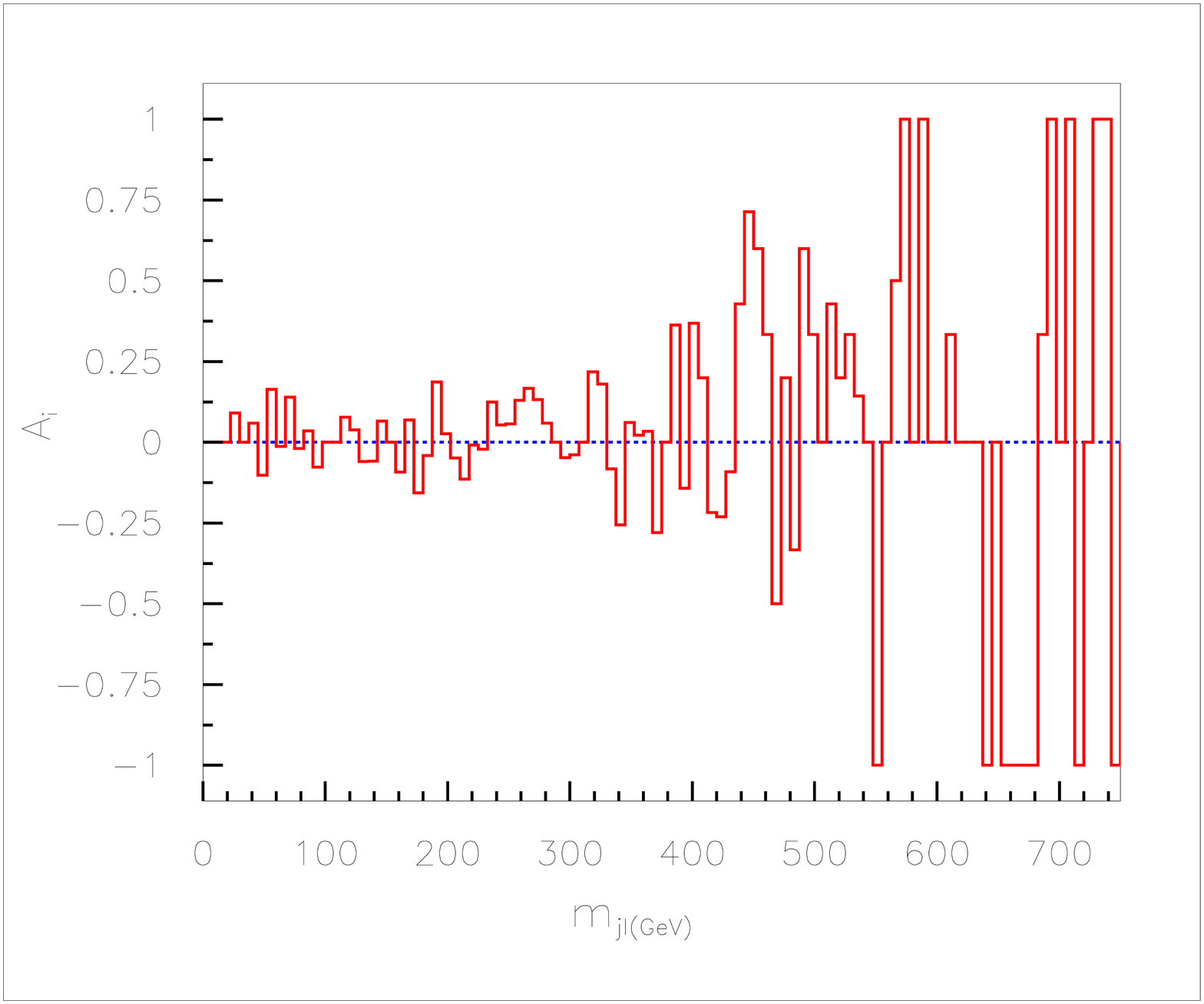,width=7.0cm,height=6.0cm,angle=-0}}
\vskip 15pt
\vskip 10pt
\centerline{\epsfig{file=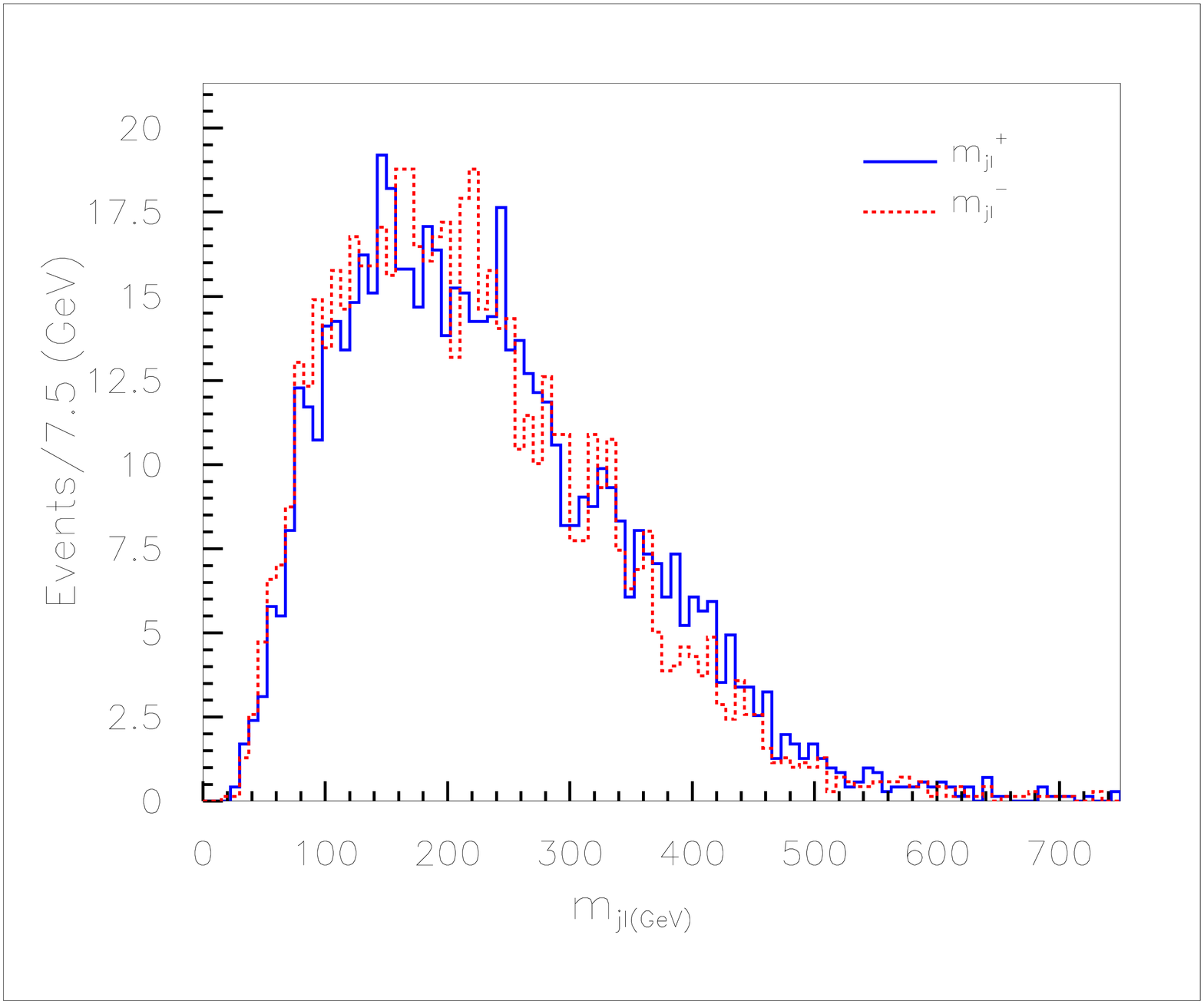,width=7.0cm,height=6.0cm,angle=-0}
\hskip 20pt \epsfig{file=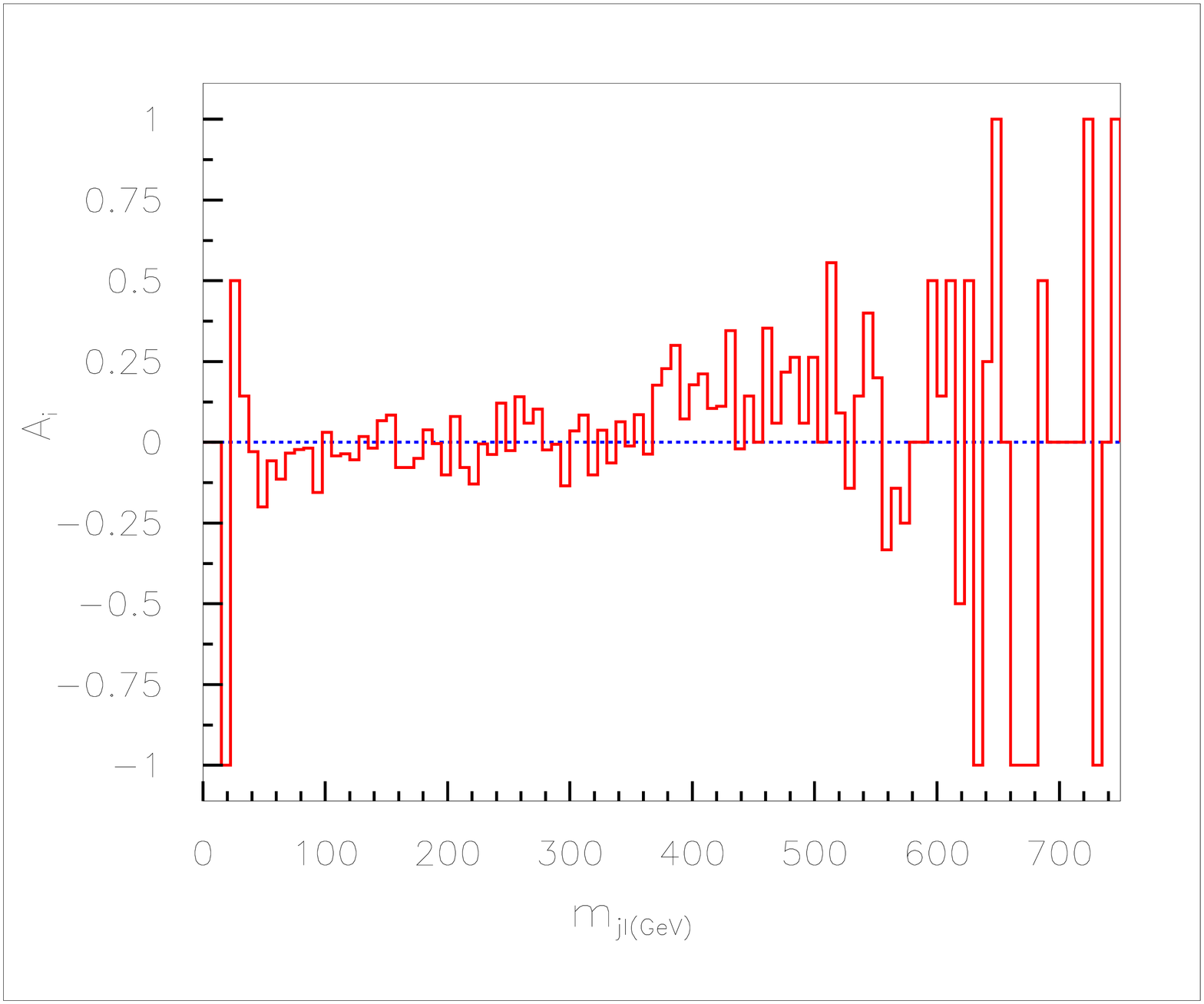,width=7.0cm,height=6.0cm,angle=-0}}
\vskip 20pt
\caption{\small \it {a) $m_{jl}$ and b) $A_i$ vs $m_{jl}$ distribution 
for mSUGRA points. The mSUGRA-1 (top) and mSUGRA-3 (bottom)
corresponds to NUHM-1,and NUHM-3 benchmark points, respectively. The event rates
are predicted at an integrated luminosity of $10 ~fb^{-1}$.}} 
\end{figure}

It is clear from Figure 3a and 3b that both for NUHM-1 and NUHM-3 we get
the desired charge asymmetry (which is negative for increasing $m_{jl}$, 
since the lighter sleptons are dominantly
left-chiral, and the leptons produced in $\chi^0_2$ decay are mostly
back-to-back with the quark while the antileptons are in the same
direction to that of the quark, hence $m_{jl^-}$ distribution has
larger population than $m_{jl^+}$ distribution near the end-point of
$m_{jl}$ invariant mass distribution.

The situation is somewhat more complicated for the corresponding mSUGRA
points. For mSUGRA-1, the sleptons are heavier than the second lightest
neutralino and the decay $\chi^0_2\r\tilde{l}^{\pm}l^{\mp}$ is
suppressed. Here the main source of the flavor subtracted opposite
sign same flavor dilepton signal is the two step $\tilde{q}_L\r
q\chi^0_2\r ql^{\pm}l^{\mp}\chi^0_1$ decay chain rather than the three
step decay chain considered earlier. In this case $\chi^0_2$ decays
into a $l^{\pm}l^{\mp}\chi^0_1$ pair via an off-shell slepton or $Z$.
The sleptons are lighter than the second lightest neutralino and mostly
dominated by the right-chiral component in mSUGRA-3. $\chi^0_2$ follows
its usual three step decay chain. The expected positive charge asymmetry 
is visible for both the mSUGRA-1 and mSUGRA-3 BP's.




\section{Summary and conclusion}

We have attempted a differentiation between mSUGRA and a scenario with
non-universal Higgs masses. The extreme situation of large negative $S$,
for which the characteristic features of the NUHM spectrum  are most prominent,
has been selected for this purpose, including three possible hierarchies
among the masses of the lightest neutralino, the lighter stau and the
tau-sneutrino. The primary channel of investigation being tau-rich, regions
in the parameter spaces of both the scenarios, giving rise to similar
ditau event rates, have been pitted against each other.

In the same-sign ditau channel, we find that the ratio defined 
as $R$, the fraction of the energy carried by the charged pion in a jet
produced in one-prong tau-decays, is a rather useful differentiator.
Because of the dependence of $R$ on the polarisation of the tau,
one ends up having different numbers of events for the two cases 
in the regions $R<0.8$ and $R>0.8$. The ratios of these two event numbers,
in turn, display a concentration in different regions, depending on whether
it is NUHM or mSUGRA.

We have further suggested the utilisation of signals involving leptons of
the first two families, which are largely left chiral in NUHM. A bin-by-bin
analysis of the of lepton-jet invariant masses exhibits a difference between
the cases with negatively and positively charged leptons, whose general
nature helps one distinguishing an NUHM scenario.

If SUSY is indeed discovered at the LHC, one will certainly  wish to 
run the machine with large integrated luminosity, so as to reveal the nature
of the underlying scenario. One important question to ask in this context 
will be whether Higgs mass(es) have different high-scale origins compared
to masses of the remaining scalars, namely, squarks and sleptons. A study
in the line suggested here, based on the polarization study of tau as
well as the first two family leptons, can be helpful in finding
an answer to such a question. 

\vspace{0.2 cm}
\noindent
{\large {\bf Acknowledgment:}}
 
\vspace{0.2 cm}
SB would like to thank Mihoko M. Nojiri and Theory Group, KEK and the Institute of Physics 
and Mathematics of the Universe for their hospitality 
while part of this work was being carried out. We also thank Nabanita Bhattacharya for her help
during the preparation of this manuscript, and Atri Bhattacharya for computational assistance.
 This work was partially supported by funding available 
from the Department of Atomic Energy, Government of India for the Regional Centre for Accelerator-
based Particle Physics, Harish-Chandra Research Institute. Computational work for this study was
partially carried out at the cluster computing facilities of KEK, Theory center and Harish-Chandra 
Research Institute ({\tt http:/$\!$/cluster.mri.ernet.in}).

\end{document}